# Visual Area of Interests based Multimodal Trajectory Prediction for Probabilistic Risk Assessment

Qiang Zhang, Lingfang Yang, Xiaoliang Zhang, Xiaolin Song, Zhi Huang

***Abstract*—Accurate and reliable prediction of driving intentions and future trajectories contributes to cooperation between human drivers and ADAS in complex traffic environments. This paper proposes a visual AOI (Area of Interest) based multimodal trajectory prediction model for probabilistic risk assessment at intersections. In this study, we find that the visual AOI implies the driving intention and is about 0.6-2.1 s ahead of the operation. Therefore, we designed a trajectory prediction model that integrates the driving intention (DI) and the multimodal trajectory (MT) predictions. The DI model was pre-trained independently to extract the driving intention using features including the visual AOI, historical vehicle states, and environmental context. The intention prediction experiments verify that the visual AOI-based DI model predicts steering intention 0.925 s ahead of the actual steering operation. The trained DI model is then integrated into the trajectory prediction model to filter multimodal trajectories. The trajectory prediction experiments show that the proposed model outperforms the state-of-the-art models. Risk assessment for traffics at intersections verifies that the proposed method achieves high accuracy and a low false alarm rate, and identifies the potential risk about 3 s before a conflict occurs.**

## I. Introduction

### A. Motivation

Although road fatalities have dropped by 1.7% from 2019 to 2021 across China, over 60,000 people lost their lives, and over two million were seriously injured yearly [1]. Most traffic accidents are caused by issues involving human drivers, such as cognitive overload, misjudgment and misoperation [2], [3]. Especially at intersections, traffic accidents are more prone to occur due to complex traffic [4].

The advancement of sensor technology has pushed the development of automated vehicles (AVs). AVs use machines to partially or fully replace human drivers in normal driving tasks and reduce traffic accidents due to human error [4]. However, fully automated driving has not yet achieved large-scale commercial adoption due to resistance from technical and ethical issues [5], costs, and public trust [6]. Therefore, human-machine cooperative driving with advanced driver assistance systems (ADAS), which allow human-driver and automated systems to be in the control loop of the vehicle simultaneously, has attracted considerable research attention [7]. In the current stage of human-machine cooperative driving, the control right is mainly held by human drivers [8], and assistance systems need to identify risks and intervene in driving based on human drivers' driving intentions [9], [10]. While ADAS provide drivers with assistive information/operation to improve driving safety, it still interacts passively with human drivers since accurately monitoring and understanding driver behaviors and intentions are still challenging. Early intention recognition means potential risks can be identified in advance, enough time remains for the driver and intelligent systems to take proper actions, and also possibilities for ADAS to interact with drivers actively. The eye movement feature implies the driver's driving intention in advance [11], so it could be an early indicator of driving intent. Incorporating the driver's intention into ego-vehicle trajectory prediction and risk assessment may reduce false alarms and generate warns timely.

In this paper, to address the risk assessment at intersections, we employ the eye movement feature for the intended trajectory prediction of ego vehicles and adopt a probabilistic risk assessment frame to resolve the motion uncertainties in trajectory prediction.

### B. Related Work

Drivers do not always send signals before a maneuver occurs. It is found that only 66% of lane change maneuvers use turn signals [12]. Therefore, it is necessary for ADAS to identify drivers' intentions so as not to distract drivers with false alarms. Predicting driver behavior before a maneuver occurs enables ADAS to focus on the potential traffic areas as early as possible, thereby detecting hazardous situations early. Xia [13] proposed a Human-like Lane Changing Intention Understanding Model (HLCIUM) for autonomous driving to identify the lane-changing intentions of surrounding vehicles by taking the speed change of surrounding vehicles as the lane change cue. Liu [14] trained HMMs to represent different driving intentions with field data collected from a flyover. The target vehicle's historical trajectory and surrounding traffic are considered in his model. When the surrounding traffic of the target vehicle is counted, the proposed prediction model is further improved. The above methods are based on machine learning methods. Recently, deep learning methods have shown remarkable achievements in driving intention prediction. Long short-term memory (LSTM) networks are capable of handling problems of time sequence generation. Su [15] constructed a surrounding-

*(Corresponding author: Zhi Huang).*

Qiang Zhang, Xiaoliang Zhang, Xiaolin Song, Zhi Huang are with the School of Mechanical and Vehicle Engineering, Hunan University, Changsha 410000, China (e-mail: zhangqiang@hnu.edu.cn; 2276085909@qq.com; jqysxl@hnu.edu.cn; huangzhi@hnu.edu.cn).

Lingfang Yang is with the School of Civil Engineering, Hunan University, Changsha 410000, China (e-mail: yanglf@hnu.edu.cn).

aware long short-term memory (LSTM) model to predict the lane change intention with its historical trajectory and neighbors' states. Huang [16] proposed a spatial-temporal convolutional long short-term memory (ConvLSTM) model to predict the vehicle's lateral and longitudinal driving intentions simultaneously. The proposed method considers the target vehicle's motion states and models its surrounding vehicles' influence and spatio-temporal interactions. Li [17] proposed an attention-based LSTM model for lane change behavior prediction on highways. The prediction model has two parts: a pre-determination model based on the C4.5 decision tree and Bagging ensemble learning, and an LSTM multi-step lane change prediction model with an attention mechanism. This model takes both the current states of surrounding vehicles and the target vehicle's historical trajectories as input and predicts the lane change intention 1s before the actual maneuver occurs.

The above driving intention prediction models, whether machine learning-based or deep learning-based, have one thing in common: they count not only the kinematics and dynamics of the target vehicles but also interactions with the surrounding environment. Different from implicitly inferred driving intention using interaction features, identifying drivers' intentions from their behavior seems straightforward, which could be an effective way in complex traffics. Studies have shown that driver's behavior contributes more to intention inference than traffic context and vehicle dynamics, especially the head posture and eye movement which show significant advantages in the early recognition of lane change intentions [11], [18]. Researchers used driver behavior to infer potential driving intentions. Xing [19] proposed an ensemble bi-directional recurrent neural network (RNN) model with Long Short-Term Memory (LSTM) units to predict lane change intention on the highway by employing the driver's head posture and eye movement feature. The results showed that with the employment of drivers' behavior feature, the model inferred the intention 0.5 seconds before operations with an average accuracy of 96.1%. Khairdoost [20] proposed a deep learning method based on long short-term memory (LSTM), which utilized the driver's movement, head position, and vehicle dynamics. The prediction model can predict three maneuvers, including left/right turns and straight ahead, and the experimental results show that the model improves the F1 score by 4% compared to the traditional IO-HMM-based intention recognition method in [21].

Accurate identification of driving intention is the cornerstone of risk assessment and is further employed for risk prediction. Li [17] utilized an attention-based LSTM model to identify lane change intentions on highways and used the estimated Time-To-Collision (TTC) to predict risks. Risk assessment for vehicles not in the same lane requires knowing their possible future positions [22], [23]. Chen [24] inferred the driver's intention using a BP neural network, and then the identified intention was used as the control matrix of a Kalman filter by which the vehicle trajectory was predicted and the collision risk probability was evaluated. Huang [25] proposes a probabilistic driving risk assessment framework based on intention recognition. In their work, an intention recognition model (IIM) is built using a long short-term memory (LSTM) network, and a driving safety field-based risk assessment model (RAM) is used to output potential risks. Shangguan [26] proposed a proactive lane-change (LC) risk prediction framework which employed a long short-term memory (LSTM) neural network with time sequences inputs to identify drivers' lane change intentions, and a light gradient boosting machine (LGBM) algorithm was used to predict driving risks. In these intention-based risk assessment methods, identified intentions act as an intermediate result connecting primitive features (such as interactive features) and the intended trajectories, while details in primitive features contributing to trajectory prediction are blocked in this framework. Wang [27] developed a two-stage multimodal prediction model for vehicle risk assessment in highway scenarios based on an intention prediction model. This model consists of a lane-change intention prediction module and a trajectory prediction module, which directly outputs multiple possible trajectories and calculates the probability of collision. This method overcomes the disadvantage of intention-based approaches. However, this multimodal trajectory prediction model only considers interactions with surroundings while ignoring the driver's behavior features.

Most studies take drivers as the dominant controller for human in-loop driving tasks. Identifying the driver's intentions would be a necessity for risk assessment. Summarizing the related work, we find that using the relatively explicit representation of driving intentions by driver behavior features has been proven effective. At the same time, the driving intention is not enough for risk assessment since it is an abstract representation of the trajectory the target vehicle will follow. Trajectory prediction could be improved by synergizing the high-level driving intention with the details in interactive features. While this issue is seldom addressed in recent work. Furthermore, the probabilistic method should be employed in trajectory prediction and risk assessment due to unmodeled errors.

### *C. Contribution*

To address the abovementioned issues, we proposed a visual AOI based multimodal trajectory prediction model and applied this model for traffic risk assessment at intersections. The prediction model adopts a parallel multimodal trajectory prediction framework in which the potential driving intention is inferred using the visual AOI feature, historical vehicle states and environmental context feature, and multimodal trajectories are predicted using the same features. Finally, the expected trajectory is predicted by combining the predicted driving intention and multimodal trajectories. A probabilistic risk assessment approach calculates the conflict probability based on the predicted trajectory and prediction error distribution. Due to the fact that eye movement data are typically unavailable in naturalistic driving datasets, such as NGSIM [28] and HighD [29], we constructed an HNU-DRE dataset (https://share.weiyun.com/oDI07QpI) consisting of vehicle dynamics, eye movement, driver's operations and video in the driver's view with a driving simulator to study the correlation between driving behavior and driving intention.

The main contributions of this paper are as follows:

1) An intention prediction model using the visual AOI feature that explicitly reflects the driver's intention. This model leverages human cognitive capabilities to handle complex traffic and provide more accurate and timely intention prediction.

2) A trajectory prediction frame by synergizing driving intention prediction with multimodal trajectory prediction. This frame combines a high-level understanding of the future maneuvers of the ego vehicle and detailed multimodal trajectories.

### D. Paper Organization

The remainder of the paper is organized as follows. Section II presents the system structure of the proposed trajectory prediction and risk assessment. The driving task-related eye movement dataset and visual AOI extraction are described in Section III. Section IV describes the details of the multi-modal trajectory prediction and the risk assessment model. Details of the experimental implementation are given in Section V. Experimental results and discussion, including ablation experiments of the intention prediction model, comparison experiments of the trajectory prediction model, and risk evaluation experiments, are given in Section VI. Section VII makes the conclusions.

## II. The Framework of Probabilistic Risk Assessment

The proposed probabilistic risk assessment framework is shown in Fig. 1, which consists of a trajectory prediction model and a risk assessment model. We incorporate a driving intention (DI) identification module into the trajectory prediction, which is employed to infer probabilities of the driving intentions using features including eye movement, environmental context, and vehicle states. Driving intentions consist of three maneuvers, i.e., going straight, right turn, and left turn. The multimodal trajectory (MT) module outputs trajectories corresponding to three driving intentions through a codec network. The trajectories filter combines the probability of driving intention from the DI module and multimodal trajectories from the MT module. It picks the trajectory with the highest probability as the expected one. The merit of this multimodal trajectory prediction model is that it generates different possible solutions for all modes to take account of the inherited multimodal space of drivers' intention or vehicle trajectories. In fact, generating only one solution may lead to an invalid trajectory since the model tries to find the average of various possible trajectories, which may not be a realistic/possible behavior of the target vehicle.

The risk assessment model utilizes the predicted trajectory and its prediction errors to evaluate future risk. The predicted trajectory is with uncertainty at each time step. Therefore, we count this uncertainty by means of collision probability. The model's prediction error statistics at each time step are first calculated, and the error boundary of the ego vehicle is determined based on the error distribution.

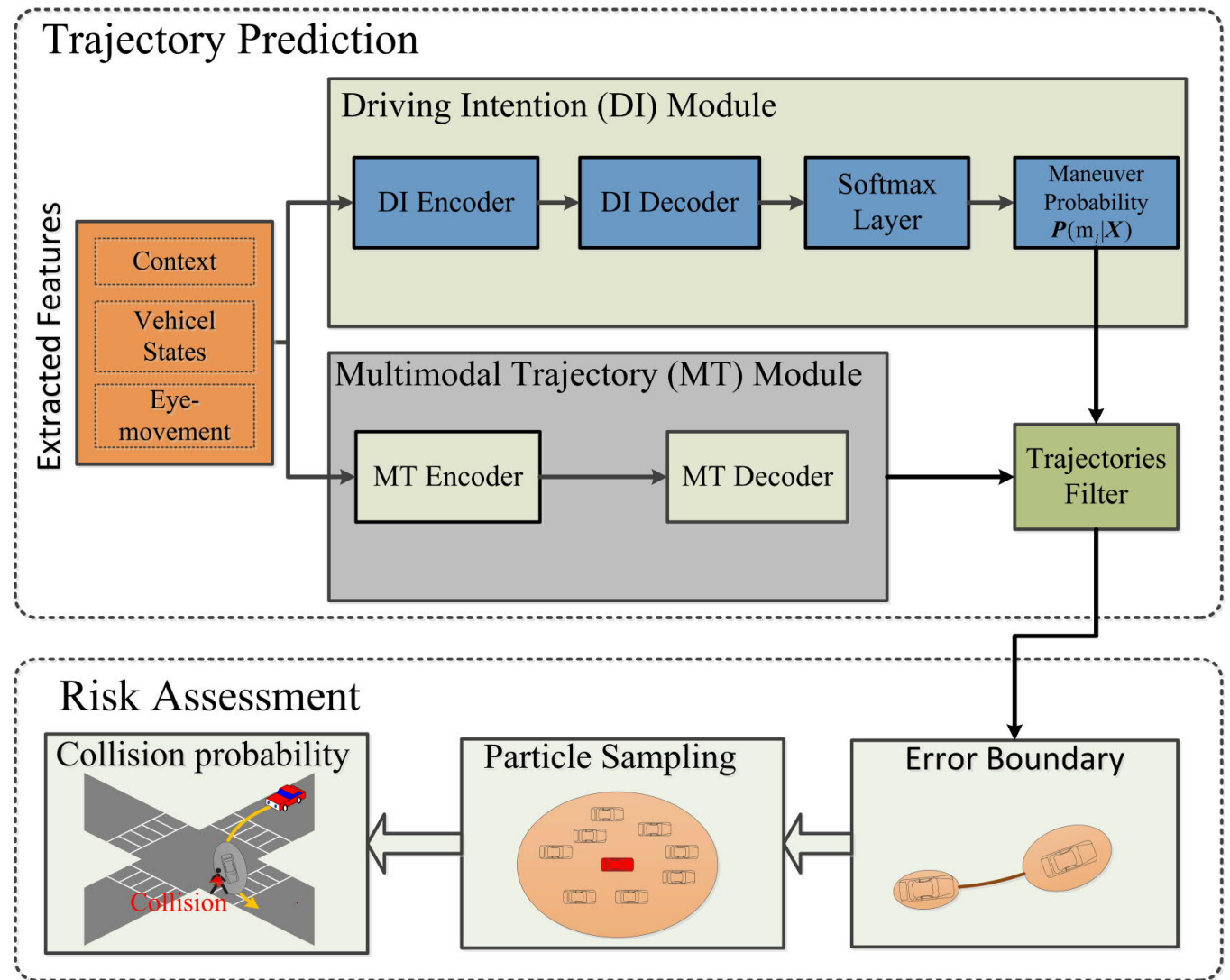

**Fig. 1.** The framework of risk assessment.

Randomly sampled particles within the error boundary are used to evaluate the collision probability between the ego vehicle and obstacles.

## III. Characteristics of Eye Movement Feature

### A. Driving Task-Related Eye Movement Dataset

There exist datasets for the study of eye movement, such as the GazeCom [30] dataset and the Andersson dataset [31], [32], which are datasets for the classification of eye movement. DR(eye)VE [33], DADA [34] and Deng [35] datasets are driving-related eye movement datasets collected with driving simulators or by watching traffic videos. Only eye movement and video are recorded in these datasets, while vehicle states are unavailable. Through driving simulation, we constructed our dataset consisting of eye movements, vehicle states, and driver's operations and videos in the driver's view. The traffic is simulated with a CARLA-based simulator which runs vehicle dynamics and physical world simulations and outputs the vehicle states and the driver's front view video with a resolution of 1920×1080. The sample rate of the video, driver's operations and vehicle states is 10 Hz. A Tobli eye tracker records the driver's eye movement in the image coordinate system at a frequency of 90Hz.

Five people with an average age of $22 \pm 3$ years were recruited to participate in eye movement experiments. They both have driver's licenses and at least one year's driving experience. All participants have normal vision and experience driving in a virtual environment with a steering wheel and pedals, e.g., playing racing games. The length of the dataset is 3422 seconds, including 109 left turns and 94 right turns. Before dataset collection, all participants practiced simulation driving for at least 30 minutes.

### B. Visual AOI Extraction

Eye movement is generally classified into fixation, saccade, smooth pursuit, or noise [30]. Eye movement behavior is active during driving due to the frequently acquiring traffic information in a wide field of view, such as rapid left-right glances at intersections and occasional checking mirrors. Therefore, not only the fixation and smooth pursuit but also the saccade is relevant to driving intention. We use the bivariate Gaussian distribution to fit the driver's visual AOI instead of classifying eye movement. Since the sample rates of eye movement and video are 90 Hz and 10 Hz, respectively, there are nine eye movement records in one video frame, as shown in Fig. 2(a). The visual AOI is described by a bi-variate Gaussian distribution, i.e., $\boldsymbol{P}_{\text{AOI}}^{(t)} \sim \mathcal{N}(\mu_x^{(t)}, \mu_y^{(t)}, \sigma_x^{(t)}, \sigma_y^{(t)}, \rho^{(t)})$, and is shown in Fig. 2(b). $\mu_x^{(t)}$ and $\mu_y^{(t)}$ are coordinates of the visual AOI center at time $t$. $\sigma_x^{(t)}, \sigma_y^{(t)}$ and $\rho^{(t)}$ are the standard deviation in horizontal direction, the standard deviation in the vertical direction, and the correlation coefficient, respectively.

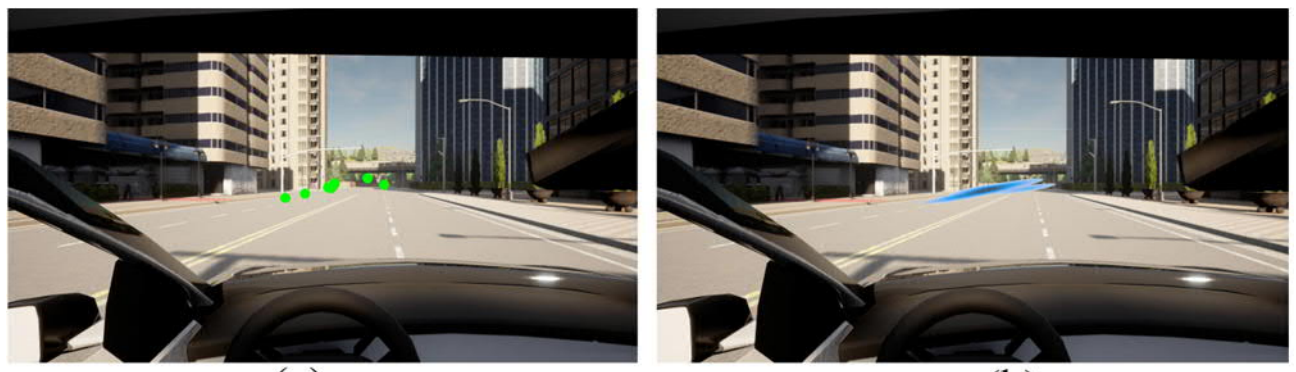

**Fig. 2.** Eye movement and its corresponding visual AOI, (a) eye movement records, (b) visual AOI.

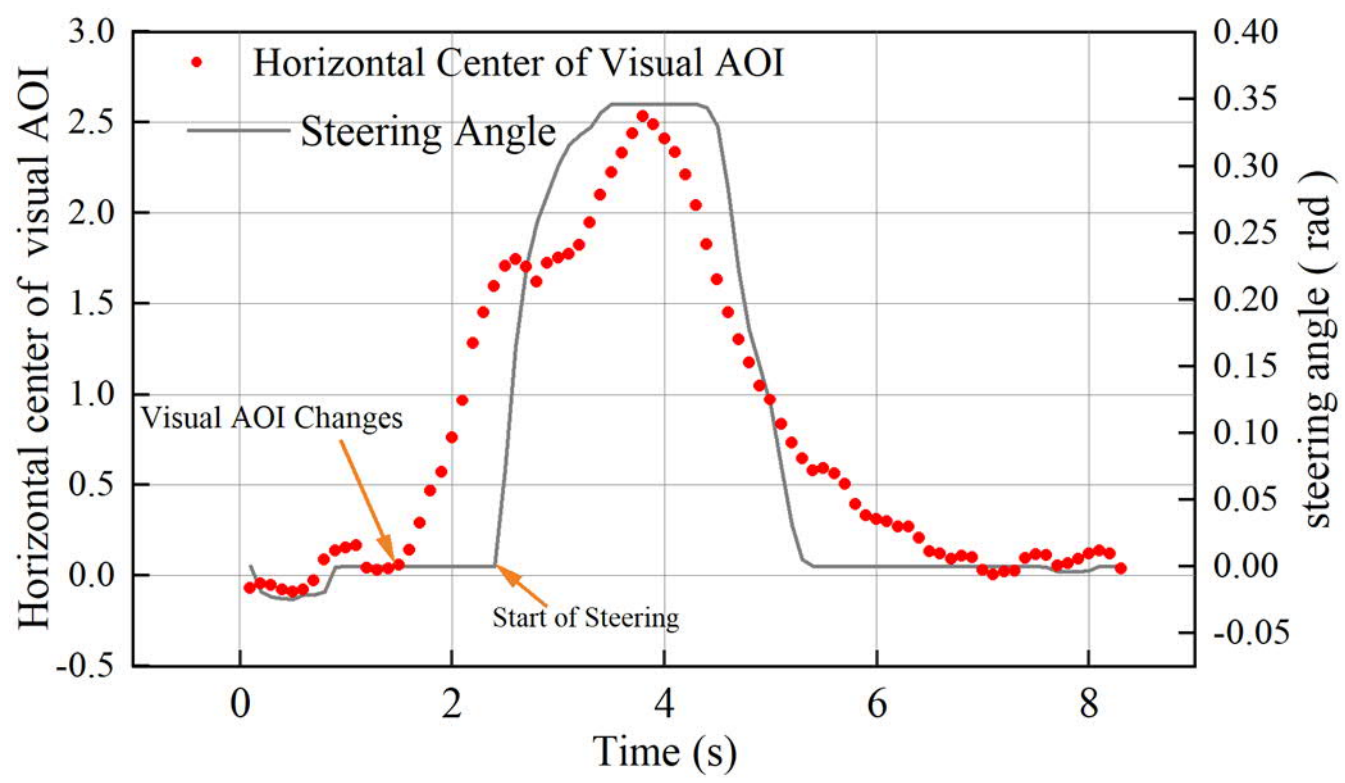

**Fig. 3.** The change of visual AOI in the horizontal direction during steering.

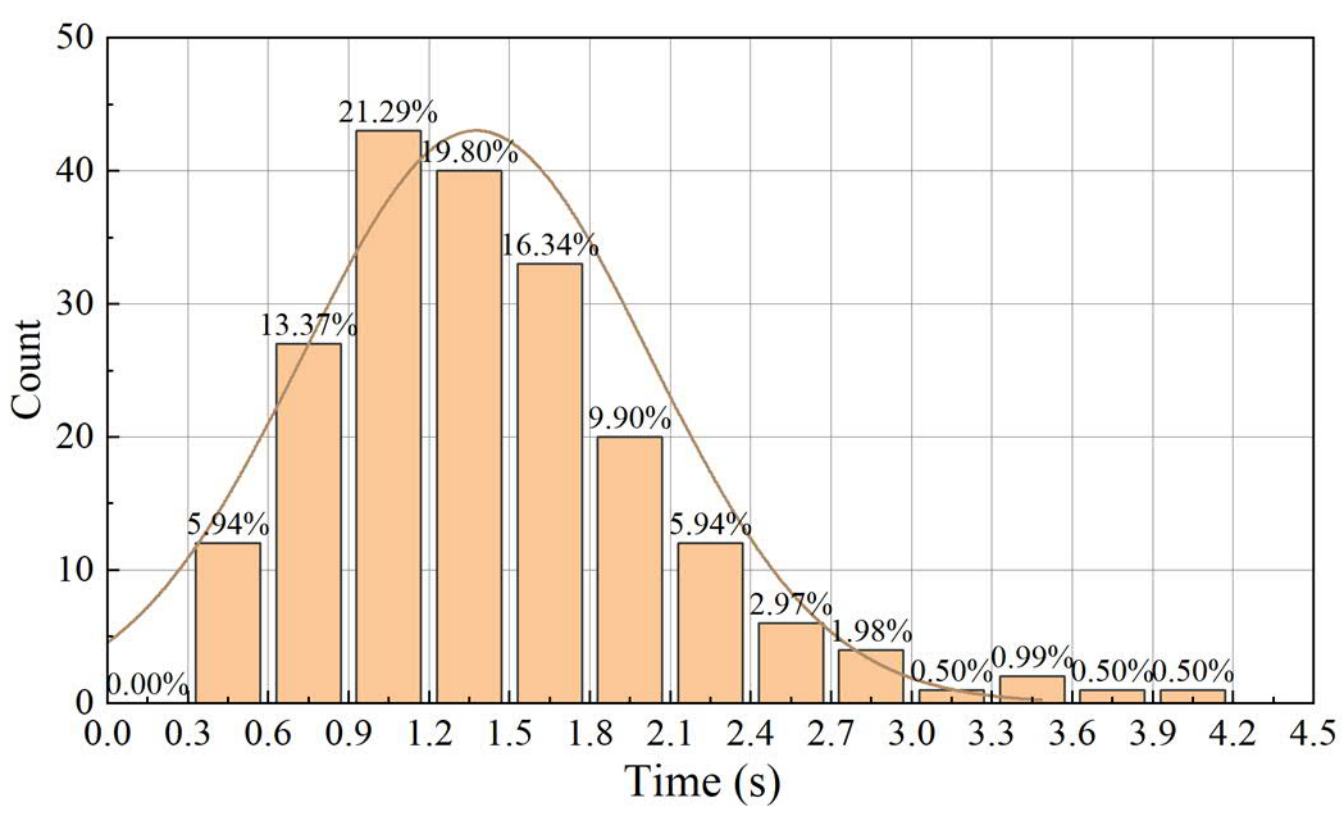

**Fig. 4.** Distribution of leading time.

### *C. Correlation Analysis of Visual AOI and Driving Operation*

Eye movement feature implies potential driving intention.

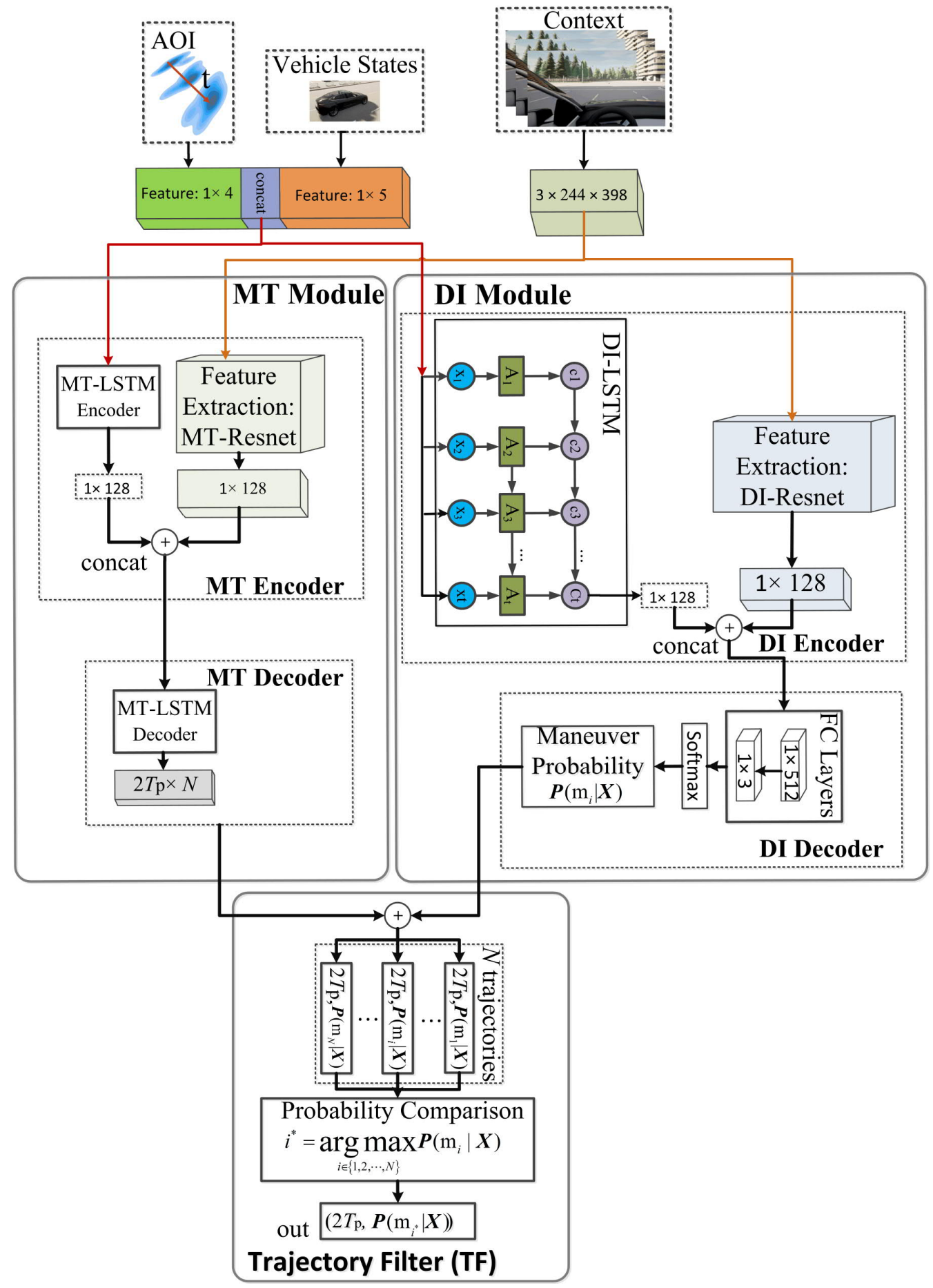

**Fig. 5.** Framework of visual AOI based trajectory prediction.

Especially in the early stage of turning, the eye movement is generally ahead of the operation, and this is manifested as an observation-action process. We quantitatively analyze the correlation between the visual AOI and driving operations. In the course of turning, eye moves distinctly in the horizontal direction. Hence, eye movement in the horizontal direction is more relevant to driving intention than the component in the vertical direction. The relationship between the AOI horizontal center $\mu_x^{(t)}$ and the steering wheel angle is studied quantitatively. An example of a typical steering process is shown in Fig. 3.

As shown in Fig. 3, a distinct correlation exists between visual AOI and steering operations. The alternation of visual AOI is ahead of the actual steering operation. We manually labeled the start time when visual AOI changes and the start time of the corresponding steering operation. The distribution of leading time ranges from 0.3 to 2.4 s and mainly concentrates in the range of 0.6-2.1 s, as shown in Fig. 4. According to the results, we can conclude that the visual AOI implies the driving intention and is about 0.6-2.1 s ahead of the operation.

## IV. Trajectory Prediction and Risk Assessment

### *A. Multimodal Trajectory Prediction Model*

The proposed trajectory prediction model is shown in Fig. 5, which consists of three parts: the driving intention (DI)

module, the multimodal trajectory (MT) module and the trajectory filter (TF). DI model predicts the probabilities of three driving intentions. Unlike the multimodal trajectory prediction model in [36], the DI model is firstly pre-trained, then its result is used to filter the output of the MT module. In the TF module, the trajectory corresponding mode with the maximum probability is selected as the predicted trajectory. As shown in Fig. 5, the driving-intention (DI) model comprises two components, i.e., the encoding module and the decoding module, denoted as DI encoder and DI decoder, respectively. The input features of the DI model are environmental context, vehicle states, and visual AOI. We use the Resnet to extract context-constrained feature from videos, such as the road feature, which is used to suppress noises in eye movement features. The vehicle states (including the vehicle's position $(p_x^{(t)}, p_y^{(t)})$, velocity $(v_x^{(t)}, v_y^{(t)})$ and heading angle $\theta^{(t)}$) are necessary for vehicle dynamics modeling and dynamics model-based trajectory prediction. DI encoder employs an LSTM (named DI-LSTM) network to encode time sequences, i.e., historical vehicle states and visual AOI. A fully connected (FC) layer and softmax function are used as the decoder to output the probabilities of driving intentions.

At intersections, there are generally three maneuvers, i.e., going straight, right turn and left turn, denoted by 0, 1 and 2, respectively. At time $t$, the vehicle states and visual AOI at time $t - t_\mathrm{h}$ to $t$ are fed into DI-LSTM for feature extraction, which is denoted as

$$\boldsymbol{X}^{(t)} = \left[\boldsymbol{x}^{(t-t_\mathrm{h})}, \cdots, \boldsymbol{x}^{(t-1)}, \boldsymbol{x}^{(t)}\right] \tag{1}$$

where $\boldsymbol{x}^{(t)} = [\boldsymbol{D}^{(t)}, \boldsymbol{E}^{(t)}]$. $\boldsymbol{D}^{(t)} = [p_x^{(t)}, p_y^{(t)}, v_x^{(t)}, v_y^{(t)}, \theta^{(t)}]$ are vehicle states described in the local coordinate system of the vehicle at time $t$, including coordinates $(p_x^{(t)}, p_y^{(t)})$, velocity components $(v_x^{(t)}, v_y^{(t)})$ in x and y direction, and heading angle $\theta^{(t)}$. $\boldsymbol{E}^{(t)} = [\mu_x^{(t)}, \mu_y^{(t)}, \sigma_x^{(t)}, \sigma_y^{(t)}]$ are Gaussian distribution parameters of the visual AOI described in the image coordinate system.

Feeding the time sequences tensor $\boldsymbol{X}^{(t)}$ into the DI-LSTM, the maneuver feature $\boldsymbol{M}_\mathrm{DI}^{(t)}$ is given by

$$\boldsymbol{M}_\mathrm{DI}^{(t)} = DI_LSTM(\boldsymbol{W}_\mathrm{DI} \cdot \left(\boldsymbol{X}^{(t)}\right) + \boldsymbol{b}_\mathrm{DI}) \tag{2}$$

where $DI_LSTM(\cdot)$ is the input-output function of the DI-LSTM, $\boldsymbol{W}_\mathrm{DI}$ and $\boldsymbol{b}_\mathrm{DI}$ are the weights and bias of DI-LSTM, respectively.

We use Resnet [37] to extract context-constrained feature from front-view videos. Videos are resized to 244×398 to reduce the computational complexity. The image at time $t$ is denoted as $\boldsymbol{I}^{(t)}$. After Resnet processing, context-constrained feature $\boldsymbol{C}_\mathrm{DI}^{(t)}$ is obtained:

$$\boldsymbol{C}_\mathrm{DI}^{(t)} = DI_Resnet\left(\boldsymbol{I}^{(t)}\right) \tag{3}$$

The dimensions of $\boldsymbol{C}_\mathrm{DI}^{(t)}$ and $\boldsymbol{I}^{(t)}$ are 128 and (3×244×398), respectively. A 256-dimensional feature is generated by concatenating $\boldsymbol{C}_\mathrm{DI}^{(t)}$ and maneuver feature $\boldsymbol{M}_\mathrm{DI}^{(t)}$, which is fed into the DI decoder to obtain the probability of the expected maneuver $\boldsymbol{P}^{(t)}(m_i|\boldsymbol{X})$:

$$\boldsymbol{P}^{(t)}(m_i|\boldsymbol{X}) = softmax\left(FC\left(concat\left(\boldsymbol{M}_\mathrm{DI}^{(t)}, \boldsymbol{C}_\mathrm{DI}^{(t)}\right)\right)\right) \tag{4}$$

where $m_i$ is the $i$-th maneuver, here $i \in (1, 2, 3)$, representing three maneuver modes. Note that the sum of the probabilities of all three modes is one.

The multimodal trajectory (MT) model also consists of two components, i.e., encoding module and decoding module, denoted as MT encoder and MT decoder, respectively. The structure and inputs of the MT encoder are the same as the DI encoder. Different from the DI decoder, the MT decoder consists of an LSTM network, and the output is a set of trajectory points from the current time $t$ to future $t + T_\mathrm{P}$. The vehicle maneuver-related feature $\boldsymbol{M}_\mathrm{MT}^{(t)}$ are extracted by MT-LSTM encoder:

$$\boldsymbol{M}_\mathrm{MT}^{(t)} = MT_Encoder(\boldsymbol{W}_\mathrm{MT} \cdot \left(\boldsymbol{X}^{(t)}\right) + \boldsymbol{b}_\mathrm{MT}) \tag{5}$$

where $MT_Encoder(\cdot)$ represents the input-output function of the MT-LSTM encoder, $\boldsymbol{W}_\mathrm{MT}$ and $\boldsymbol{b}_\mathrm{MT}$ are the weights and bias of the MT-LSTM encoder, respectively. We further extract the context-constrained feature $\boldsymbol{C}_\mathrm{MT}^{(t)}$ with the Resnet network:

$$\boldsymbol{C}_\mathrm{MT}^{(t)} = MT_Resnet(\boldsymbol{I}^{(t)}) \tag{6}$$

Concatenating the context-constrained feature $\boldsymbol{C}_\mathrm{MT}^{(t)}$ and the maneuver feature $\boldsymbol{M}_\mathrm{MT}^{(t)}$, a 256-dimensional feature is generated and fed into the MT-LSTM decoder to obtain the multimodal trajectories

$$\boldsymbol{Tra}_\mathrm{out}^{(t)} = MT_Decoder\left(concat\left(\boldsymbol{M}_\mathrm{MT}^{(t)}, \boldsymbol{C}_\mathrm{MT}^{(t)}\right)\right) \tag{7}$$

where $MT_Decoder(\cdot)$ represents the input-output function of the MT-LSTM decoder. Note that we are not to generate one trajectory but $N$ candidate trajectories, so the dimension of $\boldsymbol{Tra}_\mathrm{out}^{(t)}$ is $2T_\mathrm{P} \times N$, here $N$ = 3 corresponding to the number of maneuver modes.

Unlike the multimodal trajectory prediction model in [36] where the decoder directly outputs multiple trajectories with probabilities using a fully connected layer, the decoder in our work uses an LSTM network to output multiple trajectory waypoints at each time step.

$$\boldsymbol{Tra}_\mathrm{out}^{(t)} = \left[\boldsymbol{Tra}_1^{(t)}, \dots, \boldsymbol{Tra}_i^{(t)}, \dots, \boldsymbol{Tra}_N^{(t)}\right] \tag{8}$$

where $\boldsymbol{Tra}_i^{(t)} = [\left(p_{x_i}^{(t+1)}, p_{y_i}^{(t+1)}\right), \dots, \left(p_{x_i}^{(t+j)}, p_{y_i}^{(t+j)}\right), \dots, \left(p_{x_i}^{(t+T)}, p_{y_i}^{(t+T)}\right)]$, $\left(p_{x_i}^{(t+j)}, p_{y_i}^{(t+j)}\right)$ is the predicted waypoint of the $i$-th trajectory at the time step $t + j$, $i \in (1, 2,\dots, N)$, $j \in (1, 2,\dots, T_\mathrm{P})$, and $N$=3. In the TF module, the trajectory corresponding to the maneuver mode with the maximum probability is selected as the output.

$$i^* = \operatorname*{argmin}_{i\in\{1,\dots,N\}}\left(\boldsymbol{P}^{(t)}(m_i|\boldsymbol{X})\right) \tag{9}$$

The predicted trajectory is denoted by $\boldsymbol{Tra}_{i^*}^{(t)}$.

### *B. Risk Assessment Model*

The risk assessment model consists of three parts: prediction error boundary, particle sampling within the error boundary, and collision probability evaluation. The trajectory prediction errors are decomposed into two components: one is in the direction of the movement, and the other is in the perpendicular direction. The covariance matrix $\Sigma^{(t)}$ at time step $t$ in the prediction horizon is given by

$$\Sigma^{(t)} = E\left(\left(\mathbf{x_e}^{(t)} - \hat{\mathbf{x}}_\mathbf{e}^{(t)}\right)\left(\mathbf{x_e}^{(t)} - \hat{\mathbf{x}}_\mathbf{e}^{(t)}\right)^T\right) \tag{10}$$

where $\mathbf{x_e}^{(t)}$ is $\left[x_e^{(t)}, y_e^{(t)}\right]^T$, $x_e^{(t)}$ and $y_e^{(t)}$ are error components in the movement direction $x$ and perpendicular to the movement direction $y$, respectively, and $\hat{\mathbf{x}}_\mathbf{e}^{(t)}$ is the error mean. The error boundary with 95% confidence is represented by the joint probability density function (pdf) of the errors in the $x$ and $y$ directions.

$$P\left(\boldsymbol{X}_\mathrm{e}^{(t)}\right) = \frac{1}{2\pi\sqrt{\det\Sigma}} e^{-1/2\left(\left(\mathbf{x_e}^{(t)} - \hat{\mathbf{x}}_\mathbf{e}^{(t)}\right)^T \Sigma^{-1}\left(\mathbf{x_e}^{(t)} - \hat{\mathbf{x}}_\mathbf{e}^{(t)}\right)\right)} \tag{11}$$

For each prediction time step, we randomly generate $N_\mathrm{p}$ particles according to the error pdf within the error boundary. Each particle generates a rectangle representing the vehicle's outline. When the vehicle and obstacle share the same space, a collision occurs, and this collision event is counted. The collision probability $P_\mathrm{c}$ is obtained by dividing the number of collisions by $N_\mathrm{p}$.

$$P_\mathrm{c} = \frac{\text{Number of Collisions}}{N_\mathrm{p}} \times 100\% \tag{12}$$

## V. Implementation of Experiments

### A. Dataset and Experimental Setup

The HNU-DRE dataset is built with a driving simulator to train and evaluate the prediction model. We adopt the following model training setup: an observation horizon of 2 seconds and a prediction horizon of 3 seconds. The time steps for observation and prediction horizon are 0.1s and 0.3s, respectively. The larger time step adopted in the prediction horizon is to reduce the computational complexity of models. The driving intention is manually labeled according to the movements of visual AOI and the following maneuvers. The collected data are split by a sliding window of 5 seconds, and a total of 34,220 records, consisting of 20,429 going ahead, 5,520 right turns, and 8,271 left turns, are acquired. The ratio for training, validation, and testing are 3:1:1.

The prediction model is trained by Adam with a learning rate of 0.001. The size of hidden states of both LSTM encoders and decoders is 128, and the activation function is $\tan(\cdot)$ if not stated otherwise. We first trained the DI model separately and then incorporated it into the trajectory model to train the MT model. The presented model was trained with NVIDIA 3090 GPU in the Pytorch-frame.

### B Metrics of Trajectory Prediction and Baseline Comparison

ADE and FDE are frequently used to measure performance in previous trajectory prediction works [27], [38], where ADE is the average of the L2 distance between the ground truth and the predicted ones in the prediction horizon, and FDE is the average of the L2 distance between the ground truth and the predicted ones at final time-step $T_\mathrm{P}$. Although these two performance metrics have been widely used, it is hard to reflect the stability of the model. Therefore, in addition to the ADE and FDE metrics, we additionally employed the standard deviation of the error (SDE) [38] metric to evaluate the model's stability. SDE is the standard deviation of the L2 distance between the true and predicted trajectory and is defined as

$$SDE = \mathrm{sqrt}\left(\frac{1}{D * T_\mathrm{P}} \sum_{k=0}^{k=D-1} \sum_{t=1}^{t=T_\mathrm{P}} \left(\left\|\widehat{Tra}_k^{(t)} - Tra_k^{(t)}\right\|_2\right)^2\right) \tag{13}$$

where $\widehat{Tra}_k^{(t)}$ is the ground truth trajectory, $Tra_k^{(t)}$ is the predicted trajectory, and $D$ is the total number of records in the test dataset.

The proposed model is compared with three trajectory prediction models, i.e., CTRA, FF-LSTM, and MTP-LSTM.

- CTRA [39]: A typical vehicle motion prediction model based on constant yaw rate and acceleration assumption.
- FF-LSTM [40]: A basic encoder-decoder based on Feed Forward LSTM, where the LSTM of the decoding module takes the output from the previous time step as the input to the next time step.
- MTP-LSTM [36]: A multimodal trajectory prediction model that employs the same input features as ours except for the visual AOI. The output tensor with the dimension of $N(2T_\mathrm{p} + 1)$ is decomposed into $N$ trajectories and their corresponding probabilities by the designed MTPloss module.

### C. Potential Conflict Scenarios

We designed conflict scenarios involving vehicle-pedestrian collisions at intersections to validate the effectiveness of the proposed risk assessment framework. The conflict scenario is designed with CARLA. The traveling distance is 2440 m, including ten intersections. Drivers are required to drive on the designated road at a velocity of 10m/s and turn left or right when encountering an intersection, as shown in Fig. 6. A vehicle may collide with pedestrians in its path, and obviously, it does not collide with pedestrians not in its path.

The conflict risk is assessed at a rate of 10 Hz. The collision probabilities at each time step in the prediction horizon is evaluated, and the maximum collision probability $P_\mathrm{max}$ is taken as the predicted collision probability $P_\mathrm{c}$ at the current time, as shown in Fig. 7.

A threshold $P_s$ of 40% is adopted. When $P_\mathrm{c} > P_\mathrm{s}$, The active safety system sends warnings to driver or intervenes directly. Otherwise, there is no necessity to alert or intervene.

## VI. Results and Discussions

### A. Ablation Experiment Results of DI Model

An ablation experiment of the DI model was conducted to justify the proposed model and the input features. The experimental results are shown in Table I.

Letters S, E, C, and O represent vehicle states, visual AOI, context, and driver's operation feature. For example, S+E+C+O-LSTM represents that the vehicle states feature, visual AOI feature, context feature, and driver's operation feature are employed.

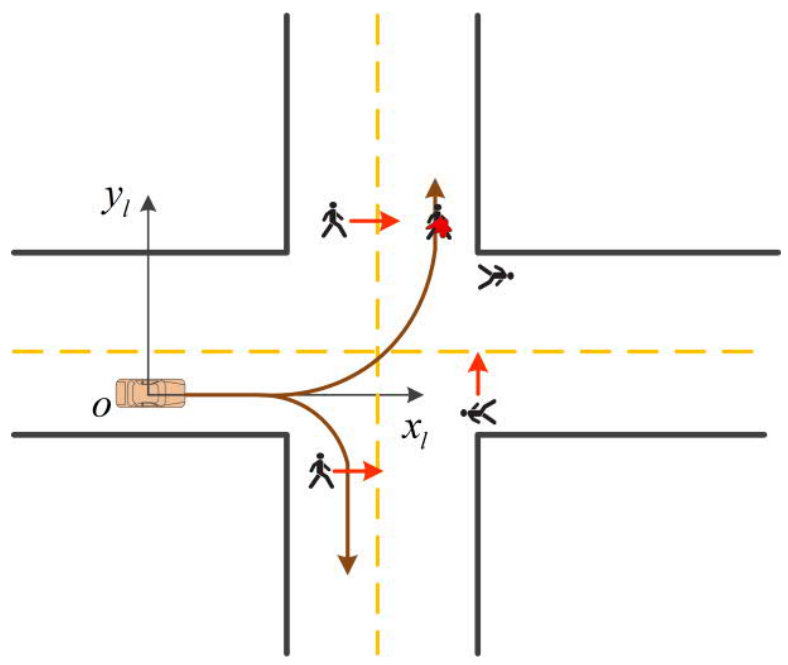

**Fig. 6.** Typical conflict scenarios for risks assessment.

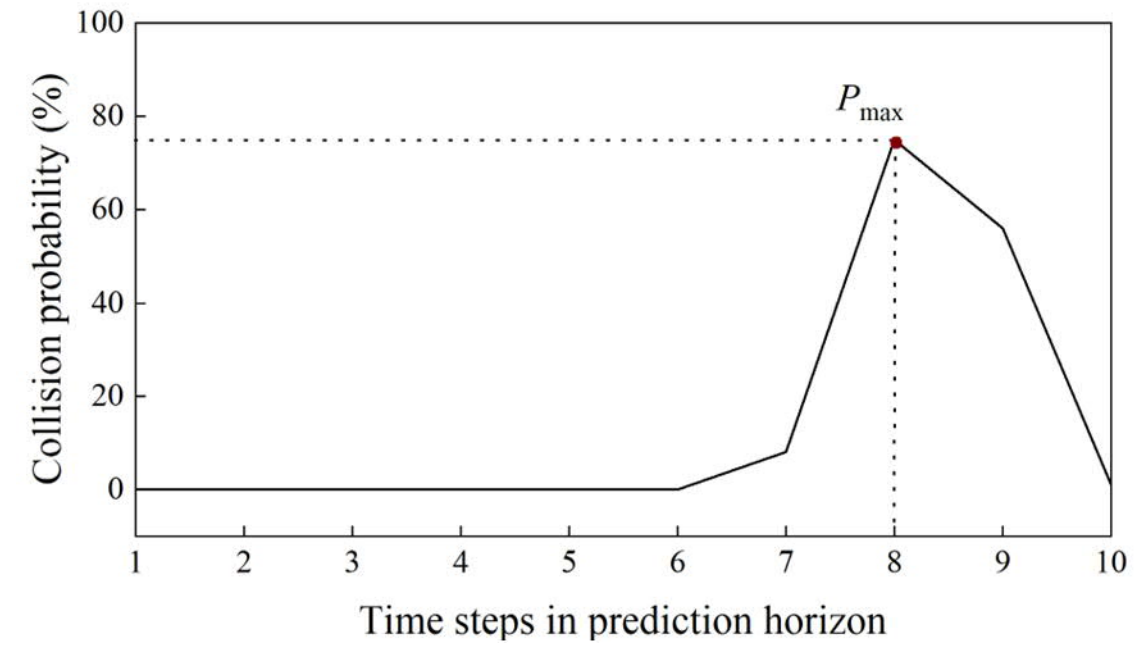

**Fig. 7.** Collision probability $P_c$ at current time.

TABLE I
RESULTS OF VARIOUS DRIVING INTENTION PREDICTION MODELS

| Models | Going Ahead | | | Right Turns | | | Left Turns | | | Time-to-Maneuver (s) |
|---|---|---|---|---|---|---|---|---|---|---|
| | Pr (%) | Re (%) | F1 | Pr (%) | Re (%) | F1 | Pr (%) | Re (%) | F1 | |
| S-LSTM | 95.46 | 93.65 | 94.55 | 92.88 | 91.90 | 92.38 | 91.45 | 95.53 | 93.44 | -0.24 |
| S+E-LSTM | 96.82 | 98.33 | 97.57 | 96.79 | 95.77 | 96.28 | 98.62 | 96.20 | 97.40 | **1.02** |
| S+E+C+O-LSTM | 97.68 | 98.55 | 98.11 | **98.22** | 97.18 | 97.69 | 98.19 | **97.09** | 97.64 | 0.832 |
| S+E+C-LSTM | **97.79** | **98.88** | **98.34** | **98.22** | **97.53** | **97.87** | **98.86** | **97.09** | **97.97** | 0.925 |

The Precision (Pr), Recall (Re), and F1 score(F1) are adopted for performance comparison. We also calculated the Time-to-Maneuver (T2M), namely the time from correctly predicting driving intention to steer operation occurring. S-LSTM is with the lowest accuracy. Incorporating the visual AOI feature improves accuracy considerably. The context feature synergizing with visual AOI further improves the accuracy. However, the synergistic enhancement effect does not happen when the operation feature is further incorporated. More features may negatively affect the model's generalization performance. The S+E+C-LSTM achieves the best accuracy. The S-LSTM only employs historical vehicle states, so its prediction lags behind the maneuver, which is verified by the negative value of T2M. On the contrary, the T2M is improved considerably when the visual AOI feature is used, which means that the visual AOI feature contributes to the earlier prediction of driving intention. We further find that with the employment of environmental context and operation features, the T2M of S+E+C+O-LSTM and S+E+C-LSTM models decreases slightly. The reason may be that operation and context features act as filters to suppress noise in the visual AOI feature. Meanwhile, the foresightedness of visual AOI is attenuated.

A case analysis further demonstrates these four models' performance differences, as shown in Fig. 8. The prediction models output 0, 1 and 2 to indicate three driving intentions, i.e., going straight, right turn, and left turn, respectively. The S+LSTM model identifies the driving intention after steering operation happens, while models using the visual AOI feature identify the driving intention slightly later than the change of the visual AOI center but ahead of the steering operation. However, if only the visual AOI feature is employed, noises in eye movement may result in false predictions, such as results of predictions at 1 s and 2 s. The context feature suppresses the influence of noises in the visual AOI feature. Based on the ablation results, we adopt the S+E+C-LSTM as the DI prediction model, denoted as EC-LSTM for simplicity. The pre-trained EC-LSTM module is then incorporated into the trajectory prediction module.

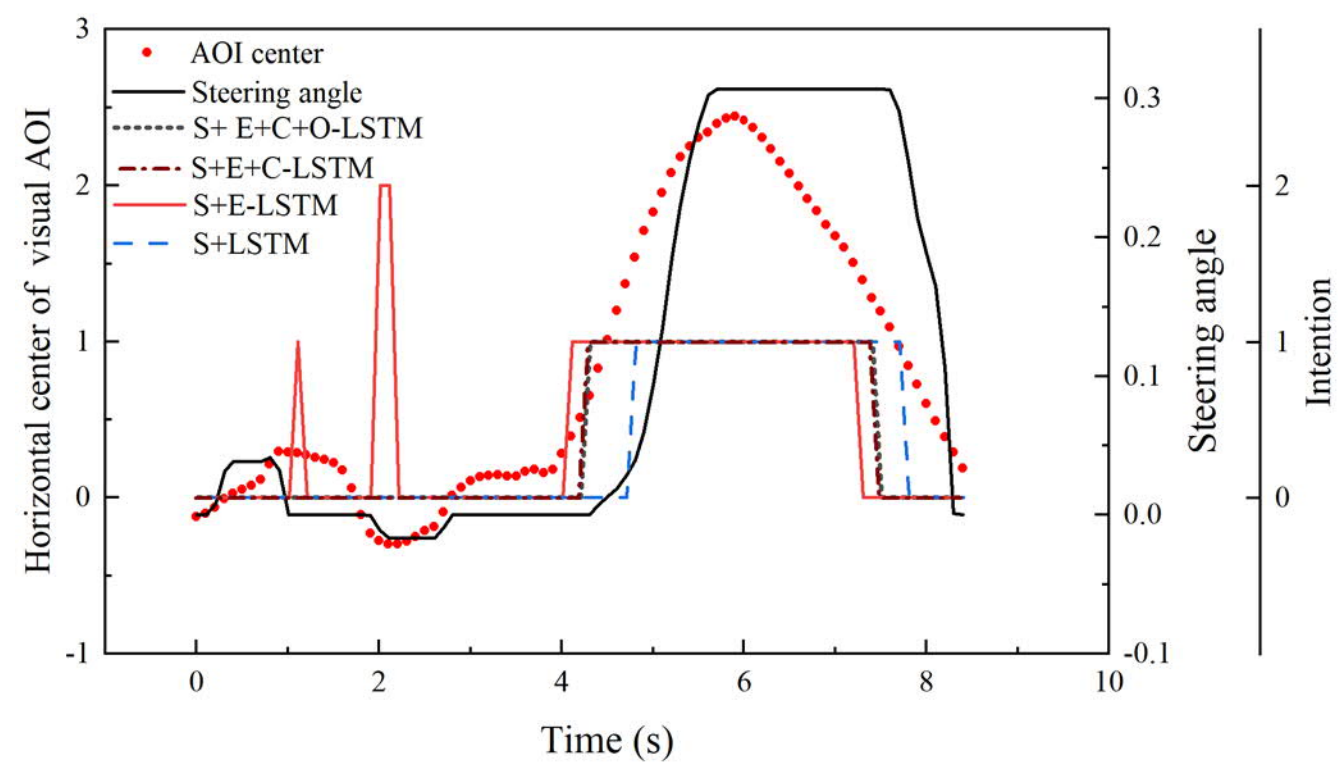

**Fig. 8.** Case analysis of driving intention prediction.

### *B. Experiment Results of Trajectory Prediction*

In the previous section, ablation experiments were done for the DI model to demonstrate the significance of the context feature and visual AOI feature for intention prediction. Similarly, we also conducted ablation experiments for the input features of the MT model. Letters C and E represent the context and visual AOI features, respectively. For example, M-EC-LSTM(C+E) indicates that both the context feature and visual AOI feature are employed in the MT model.

The results of ablation experiments and comparison experiments are shown in Table II. The prediction performances for three prediction horizons, i.e., 0.9 s, 2.1 s and 3 s, were evaluated. We can see from the ablation experiment results that M-EC-LSTM (C+E) performs best, indicating that the context feature and visual AOI feature are equally significant for trajectory prediction. Therefore, unless

TABLE II
TRAJECTORY PREDICTION RESULTS AMONG VARIOUS TRAJECTORY PREDICTION MODELS

| Models | 0.9 seconds | | | 2.1 seconds | | | 3 seconds | | |
|---|---|---|---|---|---|---|---|---|---|
| | ADE(m) | FDE(m) | SDE(m) | ADE(m) | FDE(m) | SDE(m) | ADE(m) | FDE(m) | SDE(m) |
| CTRA | **0.21** | 0.30 | 0.33 | 0.63 | 1.50 | 1.05 | 1.26 | 3.47 | 2.14 |
| MTP-LSTM | 0.33 | 0.50 | 0.28 | 0.81 | 1.51 | 0.72 | 1.15 | 2.14 | 1.03 |
| FF-LSTM | 0.34 | 0.35 | 0.30 | 0.45 | 0.70 | 0.40 | 0.66 | 1.46 | 0.75 |
| M-EC-LSTM(E) | 0.3 | 0.27 | 0.21 | 0.39 | 0.61 | 0.31 | 0.56 | 1.2 | 0.61 |
| M-EC-LSTM(C) | 0.22 | 0.25 | 0.16 | 0.33 | 0.53 | 0.30 | 0.46 | 0.93 | 0.52 |
| M-EC-LSTM(C+E) | **0.21** | **0.22** | **0.14** | **0.28** | **0.41** | **0.22** | **0.38** | **0.78** | **0.40** |

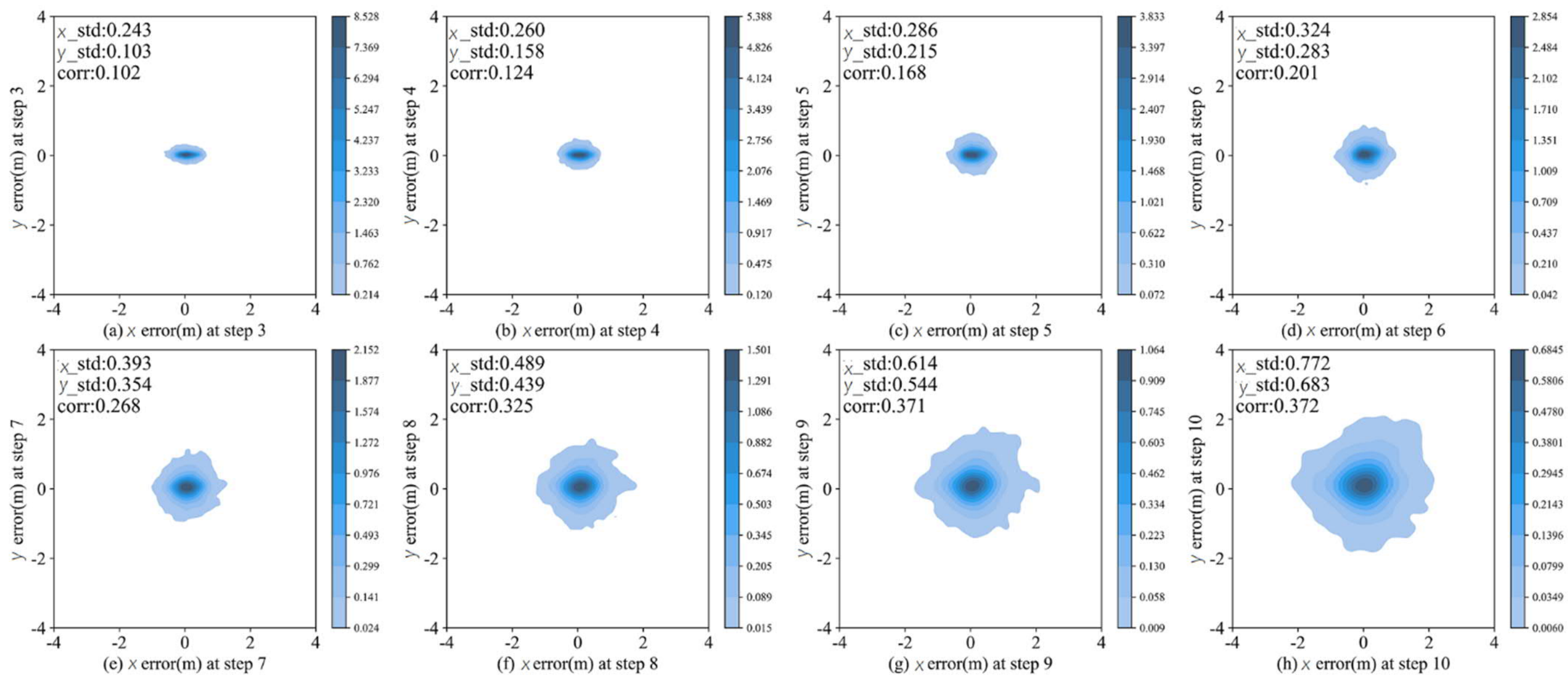

**Fig. 9.** Prediction error distribution in $x$ and $y$ directions at each time-step.

TABLE III
DISTRIBUTION OF LARGE FDE ERRORS FOR THE 3 SECONDS PREDICTION HORIZON

| Error | MTP-LSTM | FF-LSTM | M-EC-LSTM |
|---|---|---|---|
| error>1m | 37.1% | 33.1% | 19.2% |
| error>2m | 28.8% | 11.4% | 3.3% |
| error>3m | 11.2% | 4.7% | 0.8% |
| error>4m | 6.9% | 4.8% | 0.6% |

otherwise stated, we use M-EC-LSTM to represent M-EC-LSTM(C+E) for short in the rest of the paper.

Compared to other models, the M-EC-LSTM prediction model achieves the best accuracy and stability. The CTRA model surpasses the MTP-LSTM and FF-LSTM models in case of a short prediction horizon. The reason could be that the drivers generally do not change the vehicle's velocity and heading quickly. However, the CTRA model is almost inferior to all three other models for longer prediction horizon cases. No matter the short or long prediction horizons, the MTP-LSTM model works poorly. The reason may be that the visual AOI in our method helps improve the intention prediction and the probability prediction of multimodal trajectories. The FDE distributions of the 3 s prediction horizon are also given in Table III to analyze the models' performance further. It is found that M-EC-LSTM is with the smallest percentage of large errors. Significantly, the M-EC-LSTM model's distribution of errors greater than 2 m is less than 4%, much smaller than that of the other two models.

Both the predicted trajectory and the prediction error are essential for risk assessment and collision avoidance planning. We further evaluated the prediction error at each time step in the prediction horizon. The prediction errors and their components in two directions, i.e., the movement direction $x$ and the direction $y$ perpendicular to $x$, were analyzed statistically. The kernel density estimation (KDE) method was used to obtain a two-dimensional error distribution, as shown in Fig. 9. The $x$_std, $y$_std, and corr denote the standard deviation in the $x$-direction, the standard deviation in the $y$-direction, and the correlation coefficient, respectively. The standard errors in the $x$ and $y$ directions increase with the prediction time. The utilization of error distribution at each step is discussed in the following section

### *C. Experiment Results of Risk Assessment*

We devised three risk assessment methods corresponding to three trajectory prediction models, i.e., CTRA, MTP-LSTM and the proposed trajectory prediction model, denoted as CTRA-RA, MTP-RA and M-EC-RA, respectively. Moreover, according to the development practice of a collision warning system, only the conflict risk in the future 2.1 s is counted for low-speed scenarios [41].

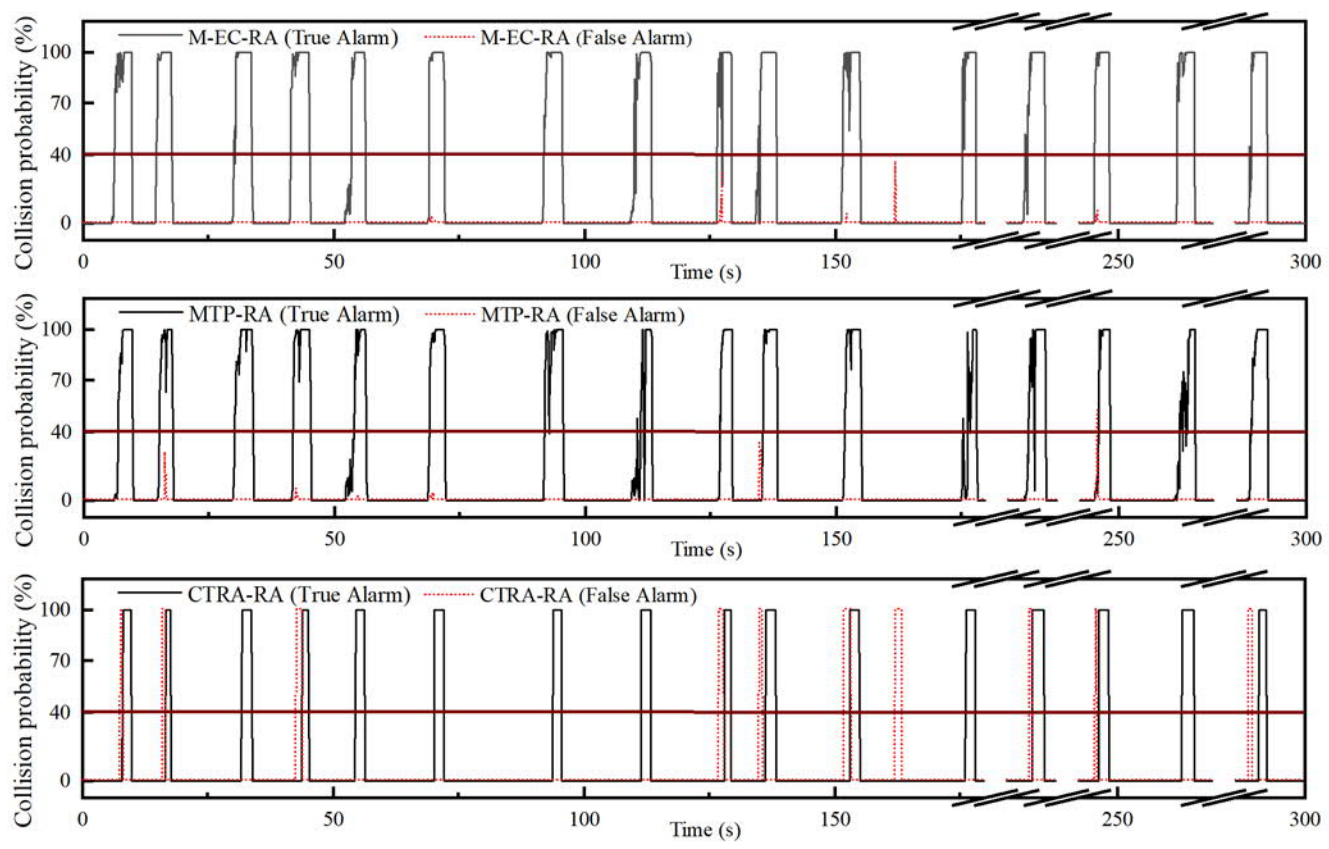

**Fig. 10.** Risk assessment results for three risk assessment algorithms.

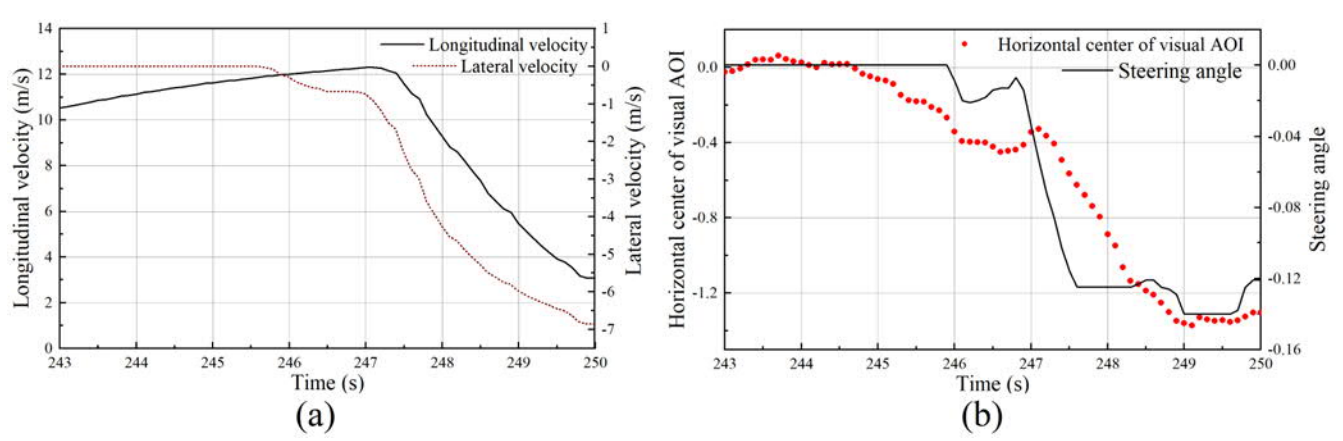

**Fig. 11.** Vehicle states, visual AOI and operation feature during a right turn, (a) the ground truth of vehicle velocity in $x_l$ and $y_l$ direction, (b) the steering angle and visual AOI center.

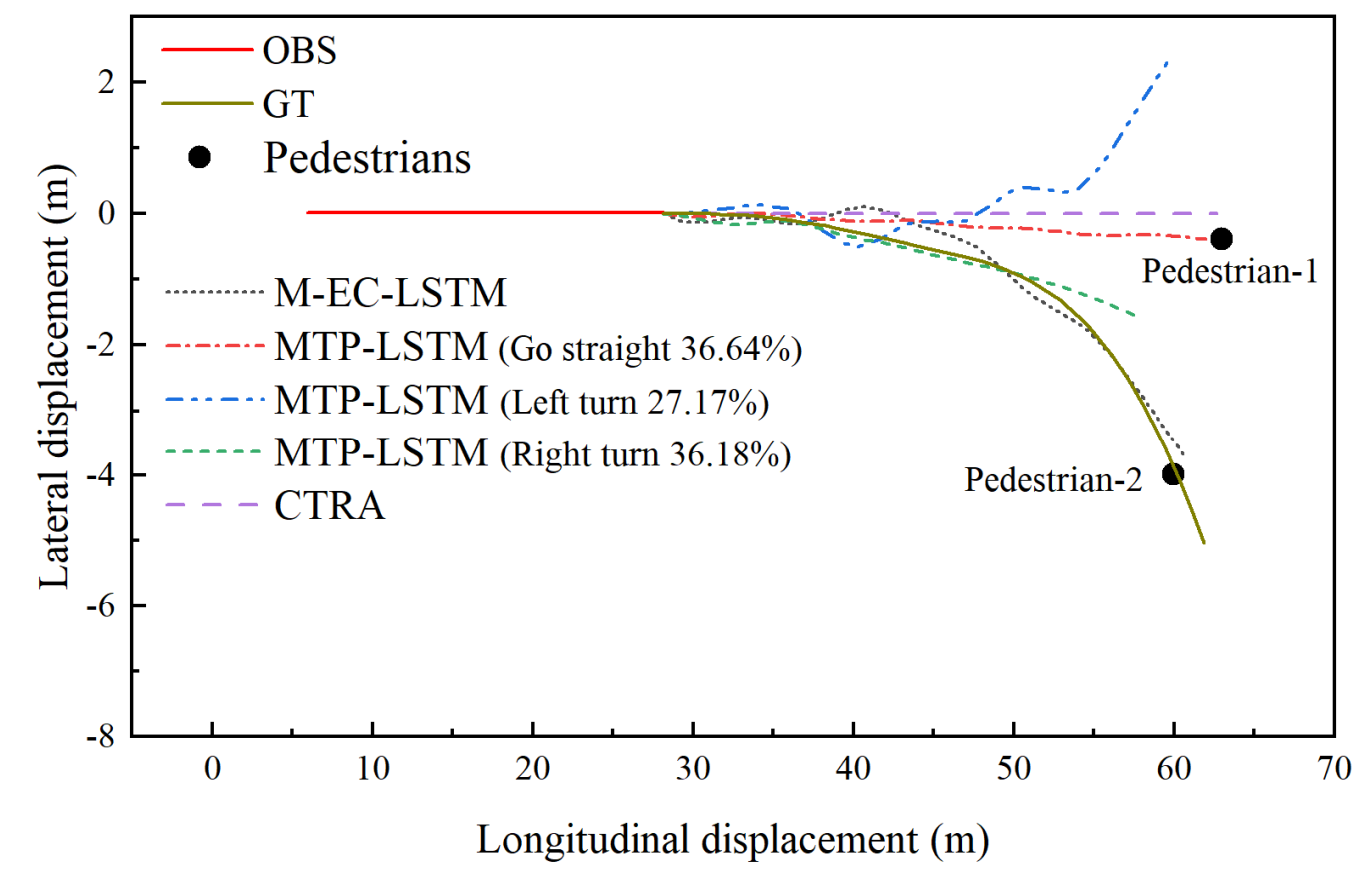

**Fig. 12.** Predicted trajectories at 245.5 s.

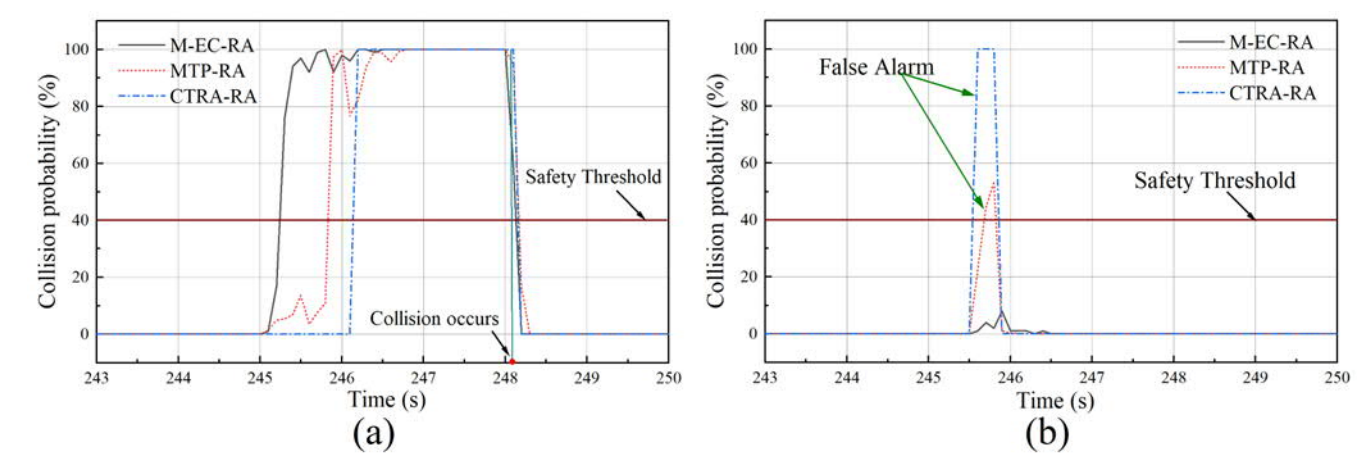

**Fig. 13.** Risk assessment during a right turn, (a) collision probability with pedestrian-2, (b) collision probability with pedestrian-1.

Since the CTRA model produces only one trajectory, CTRA-RA outputs a binary value $P_{\text{CTRA-RA}}$ indicating whether a collision with obstacles occurs on the predicted trajectory [10].

$$P_{CTRA-RA} = \begin{cases} 1, & if intersect\left(R_{ego}, R_{obs}\right) \neq \emptyset \\ 0, & otherwise. \end{cases} \tag{14}$$

where $R_{\text{ego}}$ and $R_{\text{obs}}$ describe the outlines of the ego vehicle and obstacles, respectively.

The MTP-LSTM model outputs the trajectories of three modes with the corresponding probabilities. Therefore, MTP-RA calculates the expected conflict probability of each maneuver mode using (15) [27] and takes the sum as the risk probability $P_{\text{MTP-RA}}$.

$$P_{MTP-RA} = \sum_{i=0}^{m} P\left(m_i|\boldsymbol{X}\right) P_{max}^{i} \tag{15}$$

where $P(m_i|\boldsymbol{X})$ is the probability of the $i$-th maneuver mode, and $P_{max}^{i}$ is the maximum collision probability if the predicted trajectory corresponding to the $i$-th maneuver mode is followed.

A five-minute driving simulation in the designed urban traffic was conducted to validate the proposed method's effectiveness. The driving process consists of 21 turns, sixteen involving actual collisions. The frequency of risk assessment is 10 Hz, and we set the safety threshold of the collision probability to 40%. An alarm is triggered when the collision risk is above 40% for successive three times, i.e., more than 0.2 s. The risk assessment results are shown in Fig. 10.

As can be seen from Fig. 10, all three risk assessment methods have detected all 16 true collisions. MTP-RA triggered 17 warnings, including one false alarm. The CTRA-RA triggered 26 warnings, including ten false alarms. M-EC-RA did not trigger a false alarm.

To further explore the performance differences in risk assessment, we choose a slice of right turn during 243 s-250 s for case analysis. The vehicle states, steering operation and visual AOI in the course of cornering are given in Fig. 11. We also present the predicted trajectories by the three trajectory prediction models at 245.5 s, as shown in Fig. 12. The results of the risk assessment for this right turn are shown in Fig. 13.

We can see from Fig. 11(b) that the visual AOI indicates right turn intention at around 245 s while steering operation occurs at 246 s. According to the predicted trajectory at 245.5 s, as shown in Fig. 12, the M-EC-LSTM model has identified the turning intention and generated a trajectory consistent with the steering intention. In contrast, the MTP-LSTM model predicted a right turn with a probability of 36.2% and going straight with a probability of 36.6%. Due to the lack of distinct lateral movement, the CTRA model predicted a straight forward trajectory. The predicted straight forward trajectory would result in a fake collision with pedestrian-1, as shown in Fig. 12. Thus, a false alarm might be triggered. Fig. 13(a) presents the collision probability with pedestrian-2, i.e., a true collision. Benefiting from the merit of early driving intention identification, M-EC-RA had predicted a high collision risk with pedestrian-2 3.0 s before the collision. CTRA-RA identified risk until the steering operation occurred. Although all methods satisfy the requirement of triggering true alarm 2.1 s before a collision, the advancement of M-EC-RA might be apparent in high-speed scenarios where an alarm is triggered 2.5 s even more ahead of the collision. And we can further find in Fig. 13(b) that MTP-RA and CTRA-RA triggered false alarms due to inaccurate predicted trajectories. These results verify that early identification of driving intention contributes to a low false alarm rate.

## VII. Conclusion

The paper proposes a visual AOI-based multimodal trajectory prediction model and a probabilistic risk assessment framework for complex traffic at intersections. The proposed method leverages the merits of explicit and early driving intention indication of eye movement in driving intention modeling and trajectory prediction. It accurately predicts the driving intention and the trajectory of the ego vehicle. The experiments of driving intention demonstrate that the visual AOI feature based method predicts the turning maneuver 0.925s ahead of the actual steering on average. The presented trajectory prediction model achieves smaller ADE, FDE and SDE for a prediction horizon of 3 seconds compared to other methods. Risk assessment simulation experiments in urban traffic verify that benefiting from early identification of driving intention and accurate prediction of future trajectory, the proposed method identifies collision risk accurately and timely, and the false alarm is suppressed. This study shows a perspective of applying eye movement to human-machine driving cooperation.